\pdfoutput=1 

\documentclass[12pt]{article}
\usepackage{amsmath,amssymb,amsfonts,amsthm}
\title{Nonlinear Quantum Optical Spring}

 \author{M. J. Faghihi and M. K. Tavassoly
\\
\footnotesize{Atomic and Molecular Group, Faculty  of Physics,
Yazd University, Yazd, Iran}
\\ \footnotesize{e-mail: mktavassoly@yazduni.ac.ir  } }
\usepackage{pdfpages}

\begin{document}

 \newcommand{\norm}[1]{\left\Vert#1\right\Vert}
 \newcommand{\abs}[1]{\left\vert#1\right\vert}
 \newcommand{\set}[1]{\left\{#1\right\}}
 \newcommand{\R}{\mathbb R}
 \newcommand{\I}{\mathbb{I}}
 \newcommand{\C}{\mathbb C}
 \newcommand{\eps}{\varepsilon}
 \newcommand{\To}{\longrightarrow}
 \newcommand{\BX}{\mathbf{B}(X)}
 \newcommand{\HH}{\mathfrak{H}}
 \newcommand{\D}{\mathcal{D}}
 \newcommand{\N}{\mathcal{N}}
 \newcommand{\la}{\lambda}
 \newcommand{\af}{a^{ }_F}
 \newcommand{\afd}{a^\dag_F}
 \newcommand{\afy}{a^{ }_{F^{-1}}}
 \newcommand{\afdy}{a^\dag_{F^{-1}}}
 \newcommand{\fn}{\phi^{ }_n}
 \newcommand{\HD}{\hat{\mathcal{H}}}
 \newcommand{\HDD}{\mathcal{H}}

 \maketitle
   \begin{abstract}
 The original idea of quantum optical spring arises from the requirement of quantization of the frequency
 of oscillations in Hamiltonian of harmonic oscillator. This purpose is achieved by considering
 a spring whose constant (and so its frequency) depends on the quantum states of another system.
 Recently, it is realized that by the assumption of frequency
 modulation of $\omega$ to $\omega\sqrt{1+\mu a^\dagger a}$
 the mentioned idea can be established. In the present paper we generalize the approach of quantum optical spring
 (has been called by us as nonlinear quantum optical spring) with attention to the {\it dependence of frequency
 to the intensity of radiation field} that {\it naturally} observed in nonlinear coherent states.
 Then, after the introduction of the generalized Hamiltonian of nonlinear quantum optical spring and it's solution,
 we will investigate the nonclassical properties of the obtained states. Specially, typical collapse and revival
 in the distribution functions and squeezing parameters as particular quantum features will be revealed.
 \end{abstract}

 {\bf Pacs:} 42.50.-p, 42.50.Pq, 42.50.Dv

 \section{Introduction}\label{sec-intro}
 Quantization of harmonic oscillator's Hamiltonian with the definitions
 of creation and annihilation bosnic operators is achieved. The spectrum of the system is discrete as $(n+\frac{1}{2})\hbar\omega$.
 An important point is that in quantum and classical Hamiltonians the frequency of
 oscillations ($\omega=\sqrt{k/m}$) is not quantized. The main idea of quantum optical spring lies in the
 quantization of frequency of oscillations. Recently Rai and Agarwal have designed a quantum optical spring
 such that spring constant depends on the quantum state of another system in a special form {\cite{Rai}}.
 They realized this phenomenon by replacing $\omega$
 with $\omega\sqrt{1+\mu n}$, where $n=a ^ \dag a$ is the number operator.
 Then, the Hamiltonian of quantum optical spring is introduced as
 \begin{equation}\label{Hmanko}
 H=\frac{p^{2}}{2m}+\frac{1}{2}m\omega^{2}(1+\mu n)x^{2}.
 \end{equation}
 A factor which generalizes spring constant has been called quantized source of modulation (QSM); in this case $\mu n$.
 Therefore, the eigenstate of the whole system is obtained by
 multiplying the number state with the eigenfunctions of harmonic oscillator {\cite{Rai}}.

\section{Introducing nonlinear quantum optical spring}

 It is clear that the main work of Rai and Agarwal {\cite{Rai}} can be summarized in imposing the intensity dependence of frequency of a
 quantum harmonic oscillator. Hence, in a sense we may call this
 further quantization as a second quantization  type.
 With appropriate physical motivations of the transformation has been done in Hamiltonian
 of quantum optical spring according to relation ({\ref{Hmanko}}) the authors obtained some elegant and interesting results.
 One can extend the special transformation from $\omega^2$ to $\omega^2(1+\mu n)$, to a generalized transformation
 $\omega^2(1+F(n))$, where $F(n)$ is an appropriate function of number operator.\\

 Now, recall that in the core of the nonlinear coherent states in quantum optics there exist
 an operator-valued function $f(n)$ responsible for the nonlinearity of the oscillator algebras \cite{manko1, Matos1996}.
 These states attracted much attention in recent decades \cite{sivakumar,ali}. As one of the special features of these states
 we may refer to intensity dependence of the frequency of nonclassical lights. This important aspect of
 nonlinear coherent states has been apparently clarified for the $q$-deformed coherent states with the particular
 nonlinearity function {\cite{manko2}}. The main goal of our presentation is to establish a natural link between
 "quantum optical spring" and "nonlinear coherent states" associated to nonlinear oscillator algebra, the idea that naturally
 leads one to a general formalism for {\it "nonlinear quantum optical spring"}.
 It is worth to mention that although we emphasis on the
 nonlinearity of the quantum optical spring, as we will establish
 in the continuation of the paper, the special case introduced by
 Rai and Agarwal is also nonlinear with a special nonlinearity
 function. Accordingly, our work is not a generalization from
 linear to nonlinear quantum optical spring. Instead, along finding
 the natural link between the quantum optical spring and nonlinear
 CSs, so in addition to enriching the physical basis of such systems a variety of
 quantum optical springs may be constructed.

 Single-mode nonlinear coherent states are known with deformed ladder operators $A=af(n)$ and
 $A ^\dag =f(n)a ^\dag$ where $f(n)$ is an operator-valued function responsible for
 the nonlinearity of the system.
 We assumed $f(n)$ to be real. A suitable description for the dynamics of the nonlinear oscillator is
 \begin{equation}\label{Hdeform}
 H=\omega A ^ \dag A.
 \end{equation}
 Before paying attention to the main goal of the present paper it is reasonable to have a brief
 discussion on the form of the Hamiltonian in the nonlinear
 coherent states approach. The Hamiltonian $H_M=\frac \omega 2 (A ^ \dag A + A A ^ \dag)$ has
 been introduced in \cite{manko1} in analogous to the quantized harmonic
 oscillator formalism. We have previously established in \cite{mancini} that
 requiring the "action identity" criterion on the nonlinear coherent
 states leads us to the simple form of it as expressed in (\ref{Hdeform}).
 This proposal is consistent with the ladder operators formalism and
 Hamiltonian definition have been outlined in {\it supper-symmetric
 quantum mechanics} contexts in the literature \cite{daoud,tavassoly}.
 It is worth also to
 notice that recently a general formalism for the construction of
 coherent state as eigenstate of the annihilation operator of the
 "generalized Heisenberg algebra" (GHA) is introduced
 \cite{Hassouni}. There are some physical examples there. It is easy to
 investigate that the "nonlinear coherent states" for single-mode nonlinear oscillators, as the
 algebraic generalization of standard coherent states, may be consistently placed in GHA structure
 only if one takes the Hamiltonian associated to nonlinear oscillators as in (\ref{Hdeform}). Indeed, one must consider
  $\left\{ A,      A^\dag, J_0\right\}$ with
 $J_0=H$ as the generators of the GHA. After all, our proposal allows
 us to relate simply the nonlinear coherent states to the
 one-dimensional solvable quantum systems with known discrete spectrum, i.e.,
 $f(n)=\sqrt{e_n/n}$, where $H|n\rangle=e_n|n\rangle$.

  Anyway, the time evolution operator
 \begin{equation}
 U(t)=\exp(-iH(n)t/\hbar)
 \end{equation}
 gives the following expression for the time evolved operator $A(t)$
 \begin{equation}
 A(t)=U ^ \dag(t)AU(t)=A\exp(-i\omega\Omega(n)t)
 \end{equation}
 where
 \begin{equation}\label{omega}
 \Omega(n)\equiv(n+1)f^{2}(n+1)-nf^{2}(n).
 \end{equation}
 The latter relation indicates that frequency of oscillations depends explicitly on the intensity. Now we return to the
 Hamiltonian of harmonic oscillator and deform it as follows
 \begin{equation}\label{myH}
 \HDD=\frac{p^{2}}{2m}+\frac{1}{2}m\omega^{2}\Omega^{2}(n)x^{2}
 \end{equation}
 where the term $\Omega^{2}(n)-1$  in the above Hamiltonian plays the role of QSM.
 The Hamiltonian expression in (\ref{myH}) describes the dynamics of the nonlinear quantum optical spring. Using the known solutions
 of harmonic oscillator with eigenfunction $\phi_{n}(x)$  we obtain the solutions of Schr\"{o}dinger equation
 for modulated Hamiltonian as
 $\HDD\psi^{p}_{n}\left|p\right\rangle=E^{(p)}_{n}\psi^{p}_{n}\left|p\right\rangle$
 as
 \begin{eqnarray}\label{shcrodinger}
 \psi^{p}_{n}&=& N_{n}H_{n}(\alpha_{p}x)\sqrt{\frac{\alpha_{p}}{\alpha}}\exp(-\frac{1}{2}\alpha^{2}_{p}x^{2}) \nonumber \\
 E^{p}_{n}&=&\hbar\omega\Omega(p)(n+\frac{1}{2})
 \end{eqnarray}
 where $H_n$ is the $n$th order of Hermite polynomials, $a ^ \dag a\left|p\right\rangle=p\left|p\right\rangle$,
 $\alpha_{p}\equiv(\frac{m\omega\Omega(p)}{\hbar})^{\frac{1}{2}}$, $N_{n}=(\frac{\alpha}{\sqrt{\pi}2^{n}n!})^{\frac{1}{2}}$
 and $\Omega(p)$ introduced in (\ref{omega}).
 Note that $\psi_{n}^{p}$ is the eigenstate of harmonic oscillator with frequency $\omega$ replaced by $\omega\Omega(p)$,
 and the energy eigenvalues of the modulated    Hamiltonian in (\ref{shcrodinger}) characterized  by two quantum numbers $n$ and $p$.
 For a fixed $p$ these states form a complete set. Obviously, from Eq. (\ref{omega})
 it may be seen that if $f(n)=1$ then $\Omega(n)=1$ and so $\psi^{p}_{n}$ simplifies to $\phi_{n}(x)$.
 From now on we will follow nearly similar procedure of Rai and Agarwal with the same initial state for
 the generalized modulated system we introduced
 in (\ref{Hmanko}) as
 \begin{equation}\label{inf}
 \psi(t=0)=\sum_{p,n}C_{pn}\phi_{n}(x)\left|p\right\rangle.
 \end{equation}
 Making use of the time evolution operator with the Hamiltonian in (\ref{myH}) on Eq. (\ref{inf}) we obtain
 \begin{equation}\label{evolution}
 \left|\psi(t)\right\rangle=\sum_{p,n,l}C_{pn}\exp\left(\frac{-iE^{p}_{l} t}
 {\hbar}\right)\left\langle\psi^{p}_{l}|\phi_{n}\right\rangle \left|p\right\rangle \left|\psi^{p}_{l}\right\rangle.
 \end{equation}
 For next purposes it is required to derive the density matrix with the following result
 \begin{equation}\label{density}
 \rho_{0}=\sum_{n,l,m,j,p}\left|\psi^{p}_{l}\right\rangle \left\langle\psi^{p}_{j}\right|C_{pn}C ^\ast_{pm}
 \exp\left[\frac{-i(E^{p}_{l}-E^{p}_{j}) t}{\hbar}\right]\left\langle\psi^{p}_{l}|\phi_{n}\right\rangle \left\langle \phi_{n}
 |\psi^{p}_{j}\right\rangle.
 \end{equation}
 Eq. (\ref{density}) helps us to study the quantum dynamics of the oscillator coupled to QSM. It is easy to check that setting
 \begin{equation}\label{fn-rai}
 f_{RA}(n)=\left(\frac {\sum_{j=0}^{n-1}{\sqrt{1+ \mu j}}}{n}\right)^\frac{1}{2}=\left[\frac{\sqrt{\mu}
 \left(\zeta(-\frac{1}{2},\frac{1}{\mu})-\zeta(-\frac{1}{2},n+\frac{1}{\mu})\right)}{n}\right]^\frac{1}{2}
 \end{equation}
or equivalently $\Omega_{RA}(n)=\sqrt{1+\mu n}$ in all above relations leads to the recent results of
 Rai and Agarwal in {\cite{Rai}}, where $\zeta(m,n)$ is the well-known Zeta function.
 We would like to end this section with mentioning that choosing different $f(n)$'s leads to distinct nonlinear quantum optical springs.
So our proposal can be actually considered as the generalization of their work.


\section{Quantum dynamics of the nonlinear quantum optical spring}

%
 Let us consider the situation where QSM and the oscillator are respectively prepared
 in a coherent state and in its ground state. Thus, one has
 \begin{equation}\label{inc}
    C_{pn}=\delta_{n0}\frac{\alpha^{p}\exp(-\frac{\left|\alpha\right|^{2}}{2})}{\sqrt{p!}}.
 \end{equation}
 It must be noted the equations (\ref{inc}) and (\ref{inf})
 determine the initial states. So no relation between these terms
 and the evolution Hamiltionian described the nonlinear oscillator
 may be expected.
 Inserting (\ref{inc}) into Eq. (\ref{density}) we get
 \begin{equation}\label{fdensity}
 \rho_{0}=\sum\frac{\left|\alpha\right|^{2p}\exp(-\left|\alpha\right|^{2})}{p!}\exp[-i\omega\Omega(p)t(l-j)]
 \left|\psi^{p}_{l}\right\rangle \left\langle\psi^{p}_{j}\right| \left\langle\psi^{p}_{l}|\phi_{0}\right\rangle \left\langle
 \phi_{0}|\psi^{p}_{j}\right\rangle.
 \end{equation}
 The probability of finding the oscillator in the initial state is obtained by
 \begin{equation}\label{p0}
 P_{0}(t)=\left\langle \phi_{0}\left|\rho_{0}\right|\phi_{0}\right\rangle=\sum_{p}\left|A_{p}\right|^{2}Q(p)
 \end{equation}
 where $Q(p)=\frac{\left|\alpha\right|^{2p}\exp(-\left|\alpha\right|^{2})}{p!}$ is the Poissonian distribution function and
 \begin{equation}\label{ap}
 A_{p}=\sum_{l}\exp[-i\omega\Omega(p)tl]\left| \left\langle \psi^{p}_{l}|\phi_{0}\right\rangle\right|^{2}.
 \end{equation}
 The latter formula is one of our key results for the description of nonclassical properties of the nonlinear quantum optical spring.
  Now by calculating $\left\langle \psi^{p}_{l}|\phi_{0}\right\rangle$  in (\ref{ap}) one finally arrives at
 \begin{equation}
 A_{p}=\frac{\left|\beta_{p}\right|^{2}}{\Omega(p)^{\frac{1}{2}}
 [1-(\beta^{2}_{p}-1)^{2}\exp{(-2i\omega\Omega(p)t)}]^{\frac{1}{2}}}
 \end{equation}
 where we set
 \begin{equation}
 \beta^{2}_{p}\equiv\frac{2\;\Omega(p)}{1+\Omega(p)}.
 \end{equation}
 For classical source of modulation it is enough to replace $p$ by $\left|\alpha\right|^{2}$ in Eq.
 (\ref{p0}). Consequently
 \begin{equation}\label{pcl}
 P_{cl}(t)=\frac{\left|\beta_{\alpha}\right|^{2}}{\sqrt{\Omega^{2}(\left|\alpha\right|^{2})(1-2(\beta^{2}_
 {\alpha}-1)^{2}\cos(2\omega_{\alpha}t)+(\beta^{2}_{\alpha}-1)^{4})}}
 \end{equation}
 where $\omega_{\alpha}\equiv\omega\Omega(\left|\alpha\right|^{2})$. Clearly $P_{cl}$ oscillates at frequency $2\omega_{\alpha}$.


\section{ Squeezing properties of nonlinear quantum optical spring}

%
%
 Since the number of photons and so the Hamiltonian operator in (\ref{Hdeform}) is constant, we have the following relations
 \begin{eqnarray}\label{xp}
 x(t)&=&x(0)\cos(\omega\Omega(n)t)+\frac{p(0)}{m\omega\Omega(n)}\sin(\omega\Omega(n)t) \nonumber \\
 p(t)&=&p(0)\cos(\omega\Omega(n)t)-m\omega\Omega(n)x(0)\sin(\omega\Omega(n)t).
 \end{eqnarray}
 Now we define the squeezing parameters  as
 \begin{equation}\label{sparameter}
 S_{x}(t)=\frac{\left\langle x^{2}(t)\right\rangle-\left\langle x(t)\right\rangle^{2}}{\left\langle x^{2}(0)
 \right\rangle}, \;\;\;\;\;\; S_{p}(t)=\frac{\left\langle p^{2}(t)\right\rangle-\left\langle p(t)\right\rangle^{2}}
 {\left\langle p^{2}(0)\right\rangle}.
 \end{equation}
 The expectation values must be calculated with respect to the states in (\ref{inf}). As a result it is easy to show
 \begin{eqnarray}\label{squeezing}
 S_{x}(t)&=&1-\sum_{n}\frac{\left|\alpha\right|^{2n}\exp(-\left|\alpha\right|^{2})\sin^{2}(\omega \Omega({n})t)}{n!}
 \left( 1-  \frac{1}{\Omega^{2}({n})}\right) \nonumber \\
 S_{p}(t)&=&1+\sum_{n}\frac{\left|\alpha\right|^{2n}\exp(-\left|\alpha\right|^{2})\sin^{2}(\omega \Omega({n})t)}{n!} \;
 \left(\Omega^{2}(n)-1\right).
 \end{eqnarray}
 Note that unlike the special case considered by Rai and Agarwal with $\Omega^2(n)=1+\mu n$ for which $S_{x}$ is always less
 than one ($x$-quadrature is always squeezed) and hence $p$-quadrature is not squeezed at all, that is not so in general.
 All we may conclude from the two latter relations are that squeezing in both quadratures may be occurred
  (certainly not simultaneously) depending on the selected nonlinearity function $f(n)$.


 \section{Physical applications of the formalism}\label{examples}

 Now we like to apply the presented formalism to a few classes of nonlinear coherent states. There exist
 various nonlinearity functions in the literature which have     been introduced for different purposes, mainly due to their
 nonclassical properties. Among them we only deal with two
 classes of them, i.e., "$q$-deformed coherent states" and
 "photon-added coherent states".
 \begin{itemize}
 \item{{\it $q$-deformed coherent states}}\\
 As the first example we use the $q$-deformed nonlinearity function. Recall that Man'ko {\it et al} showed
 that $q$-coherent states are indeed nonlinear coherent states with nonlinearity function \cite{manko2}
 \begin{equation}\label{fq}
 f_{q}(p)=\sqrt{\frac{q^{p}-q^{-p}}{p(q-q^{-1})}}=\sqrt{\frac{\sinh(\lambda p)}{p\sinh\lambda}}
 \end{equation}
 where $q=\ln\lambda$. The corresponding states have frequency dependence on intensity of radiation field
 (blue shift) \cite{manko2}. By replacing Eq. (\ref{fq}) into  (\ref{omega}) one readily gets
 \begin{equation}\label{omegafq}
 \Omega_{q}(p)=\frac{\cosh\left[\frac{\lambda}{2}(2p+1)\right]}{\cosh\frac{\lambda}{2}}
 \end{equation}

 \item{{\it Photon added coherent states}}\\
 As the next example we will consider photon-added coherent states
 (PACSs) first introduced by Agarwal and Tara \cite{tara} as $|\alpha,m\rangle=N (a^\dagger)^m|\alpha\rangle$,
  where $|\alpha\rangle$ is the canonical coherent states, $N$ is an appropriate normalization factor and $m$ is a non-negative integer.
   Sivakumar realized that these states satisfy the eigen-value equation
 $A|\alpha,m\rangle=\alpha|\alpha,m\rangle$ where $A=f(n,m) a$ \cite{sivakumar}. Hence, these
 states are nonlinear coherent states with nonlinearity function
 \begin{equation}\label{fs}
    f_{PACS}(n,m)=1-\frac{m}{1+n}.
\end{equation}
 Thus, by replacing (\ref{fs}) into Eq. (\ref{omega}) we have
\begin{equation}\label{omegas}
    \Omega_{PACS}(p, \; m)=(p+1)\left(\frac{p+2-m}{p+2}\right)^2-p\left(\frac{p+1-m}{p+1}\right)^2.
\end{equation}
Also, it is shown that the operator-valued function associated
with $|\alpha,-m\rangle$ is $f_{PACS}(n,-m)=1+m/(1+n)$. Replacing
$m$ by $-m$ in (\ref{omegas}) one may obtain $\Omega_{PACS}(p,-m)$
associated to $|\alpha,-m\rangle$ states \cite{sivakumar}.
 \end{itemize}
 Our numerical results have been displayed  in figures 1 and 2 show respectively the probability distribution function
 and squeezing parameter in $x$-quadrature  against $\tau$ for $q$-deformed coherent states with $\Omega_{q}(p)$ introduced
 in (\ref{omegafq}) and the choosed parameters (note that the mean photon number is indeed $|\alpha|^2$ and $\tau =
 \omega t/(2\pi)$ in all figures). A typical collapse and revival exhibition may be observed from the two figures,
 which is due to the discrete nature of the quantum state of the source of modulation (quantization of the frequency of the oscillation).
 In figures 3-a (for the states $|\alpha, m\rangle$) and 3-b, 3-c
  (for the states $|\alpha, -m\rangle$) we displayed the probability distribution as a function of $\tau$ for the different parameters.
  Also  in figure 4 (for the states $|\alpha, m\rangle$), and figures 5-a, 5-b, 5-c (for the states
  $|\alpha, -m\rangle$) the squeezing parameter in $x$-quadrature have been shown against $\tau$ for the choosed parameters.
  A typical collapse and revival exhibition can be observed from all figures, which again have their
  roots in the above mentioned source, i.e., the spring constant is controlled by another quantum source (QSM).
  From figures 4, 5-b (for the states $|\alpha, -m\rangle$)
  the squeezing exhibition may be observed, while this nonclassical feature may not be seen
  from figures 5-a and 5-c. It is worth to notice that although the two latter figures do
  not show squeezing feature,  the collapses and revivals as quantum features of the special
  type of nonlinear quantum optical spring are visible.
  To this end we would like to mention that as we stated before
  all cases considered by us and the special case of Rai and Agarwal
  quantum optical springs are nonlinear in nature according to
  our terminology, so the general features of all are the same at least
  qualitatively.

\section{Conclusion}\label{examples}
 The presented formalism in the present manuscript compare to Rai and Agarwal method has the advantage that the quantization
 of frequency of oscillations which comes out naturally from the nonlinear coherent states approach were imposed
 on the quantized Hamiltonian of harmonic oscillator.
   So in this way we impose a second quantization type on the
   quantized harmonic oscillator.
   Also our approach can be easily used for a wide range of nonlinear
   oscillators as well as every solvable quantum systems, due to the simple relation $e_{n}=nf^{2}(n)$
   \cite{our, Roknizadeh, tavassoly}. This approach leads to typical collapses and revivals in probability
   distribution function $P_{0}$. The latter results in addition to the squeezing parameter $S_{x}$ indicate
   that these are purely quantum mechanical phenomena.\\

 \begin{flushleft}
 {\bf Acknowledgements}\\
 \end{flushleft}

 One of us (M.J.F.) would like to acknowledge useful discussions with G. R.
 Honarasaand O. Safaeian regarding nonlinear coherent states.


\includepdf[pages={1}]{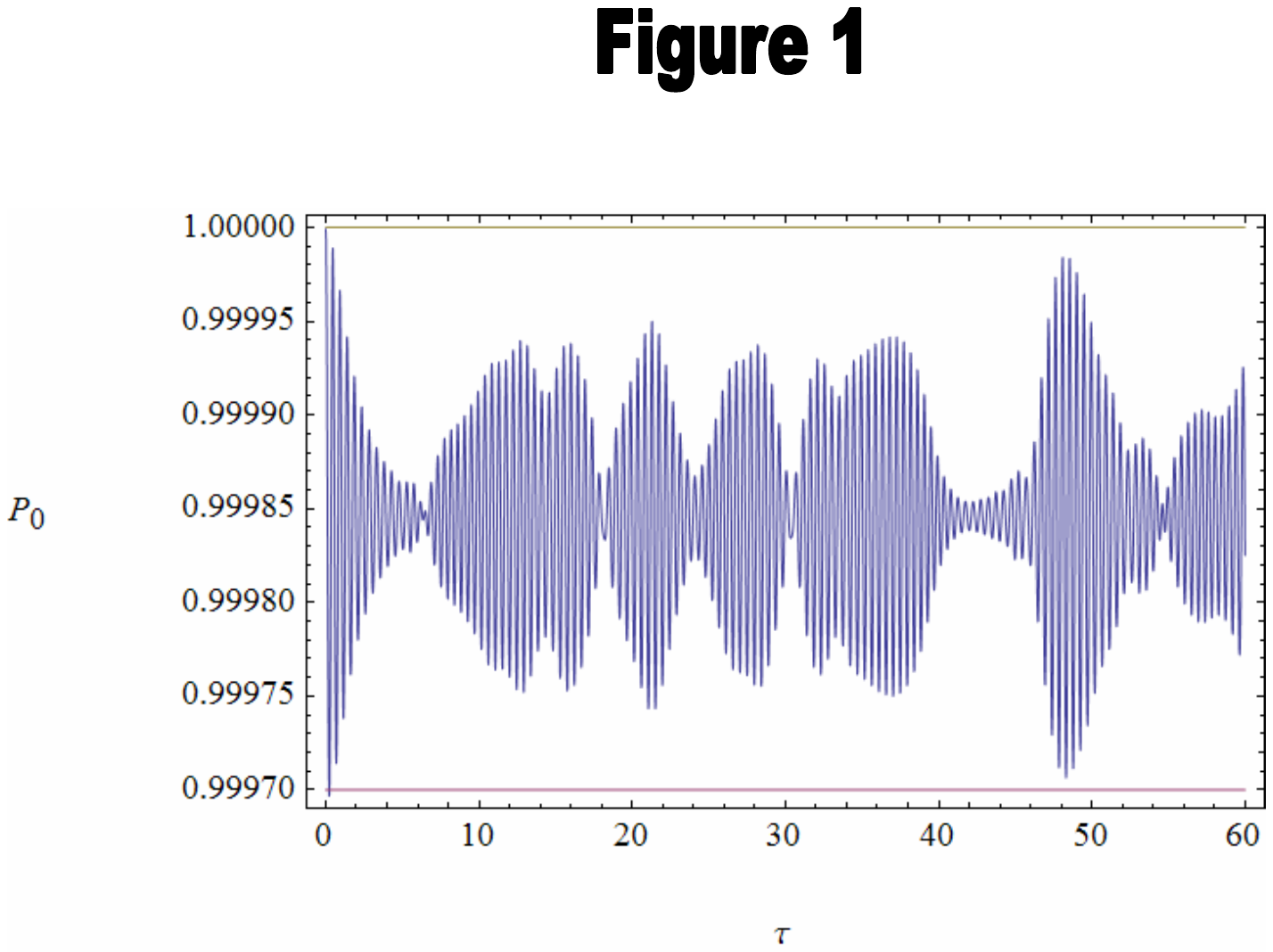}
\includepdf[pages={1}]{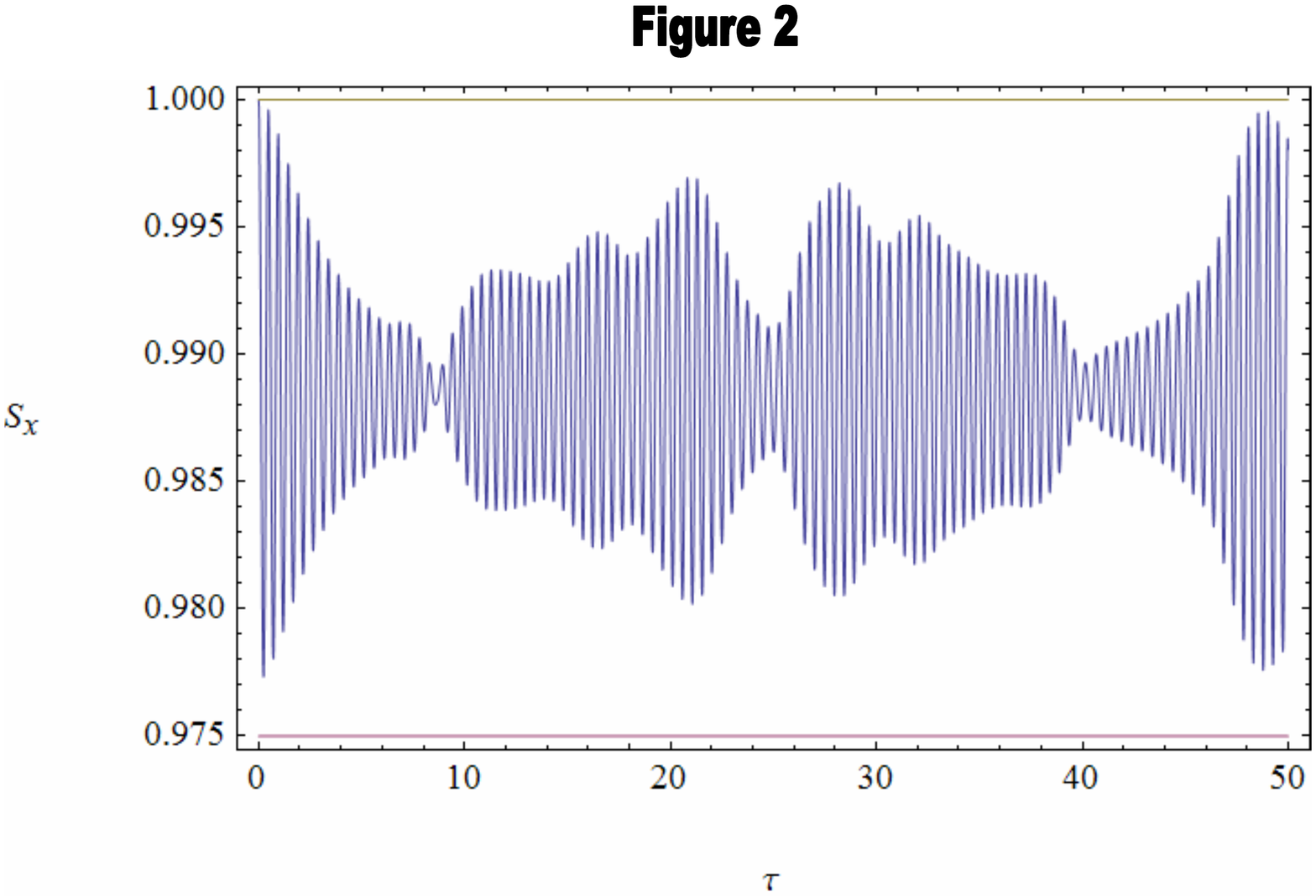}
\includepdf[pages={1}]{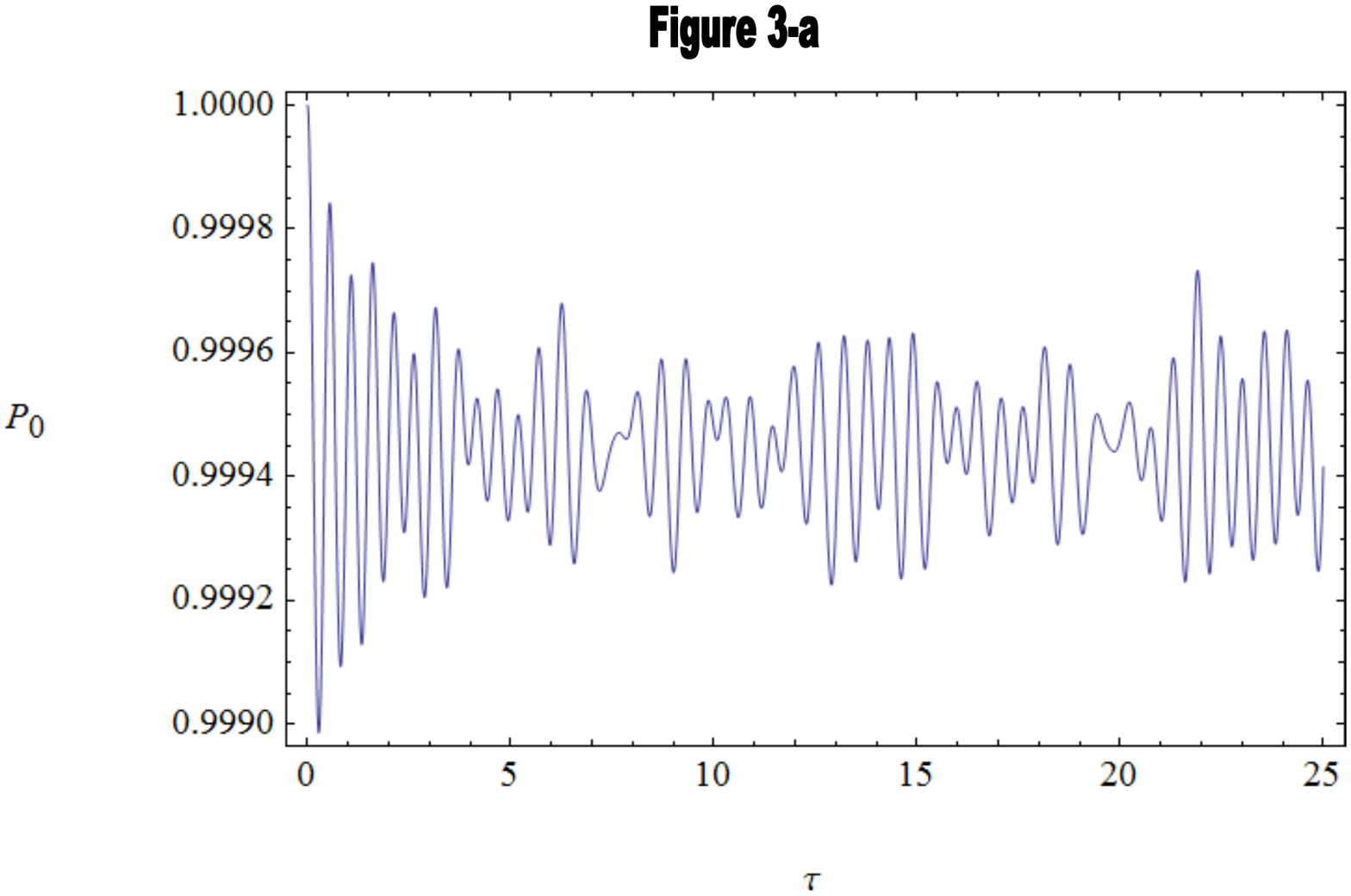}
\includepdf[pages={1}]{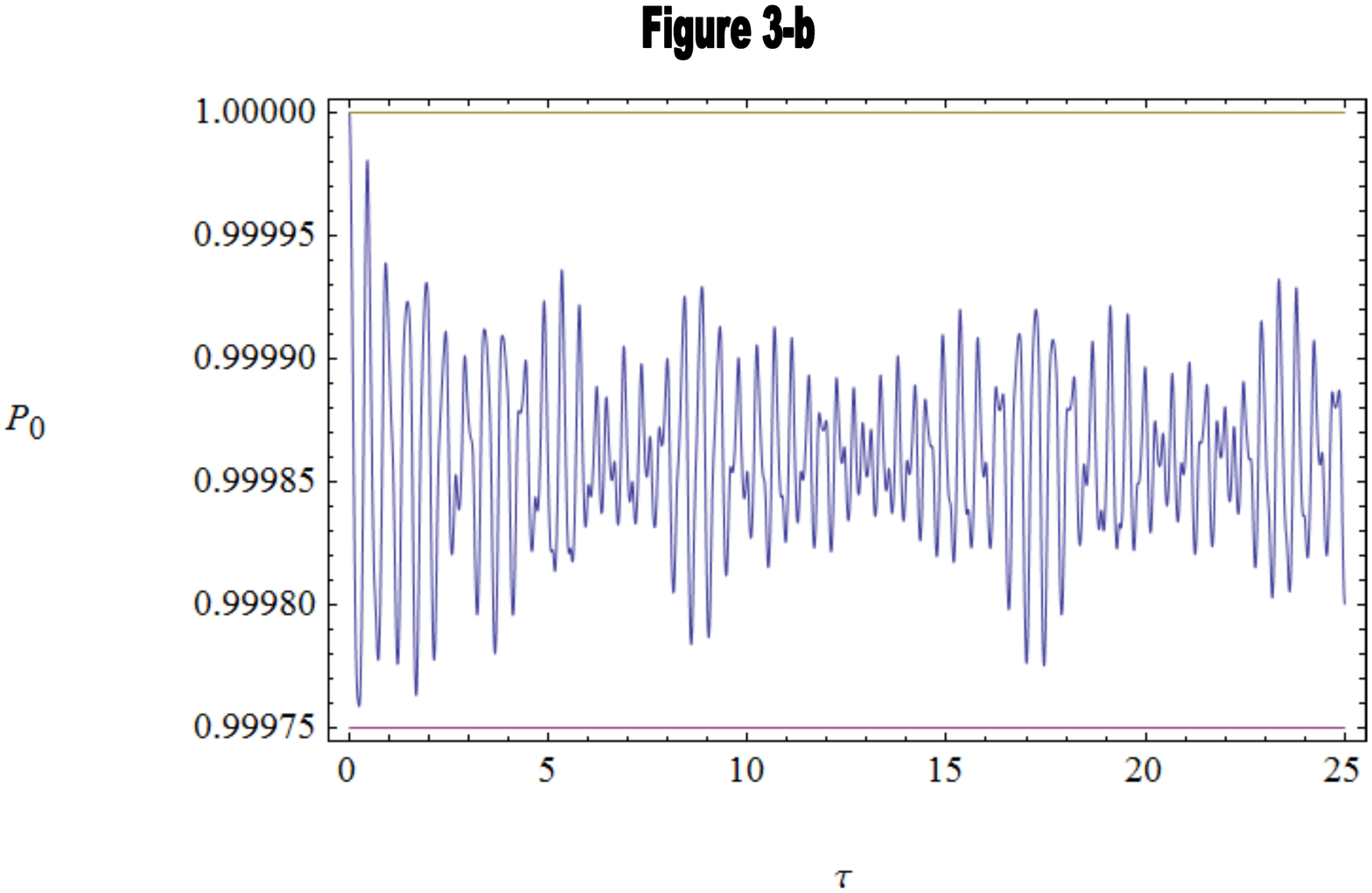}
\includepdf[pages={1}]{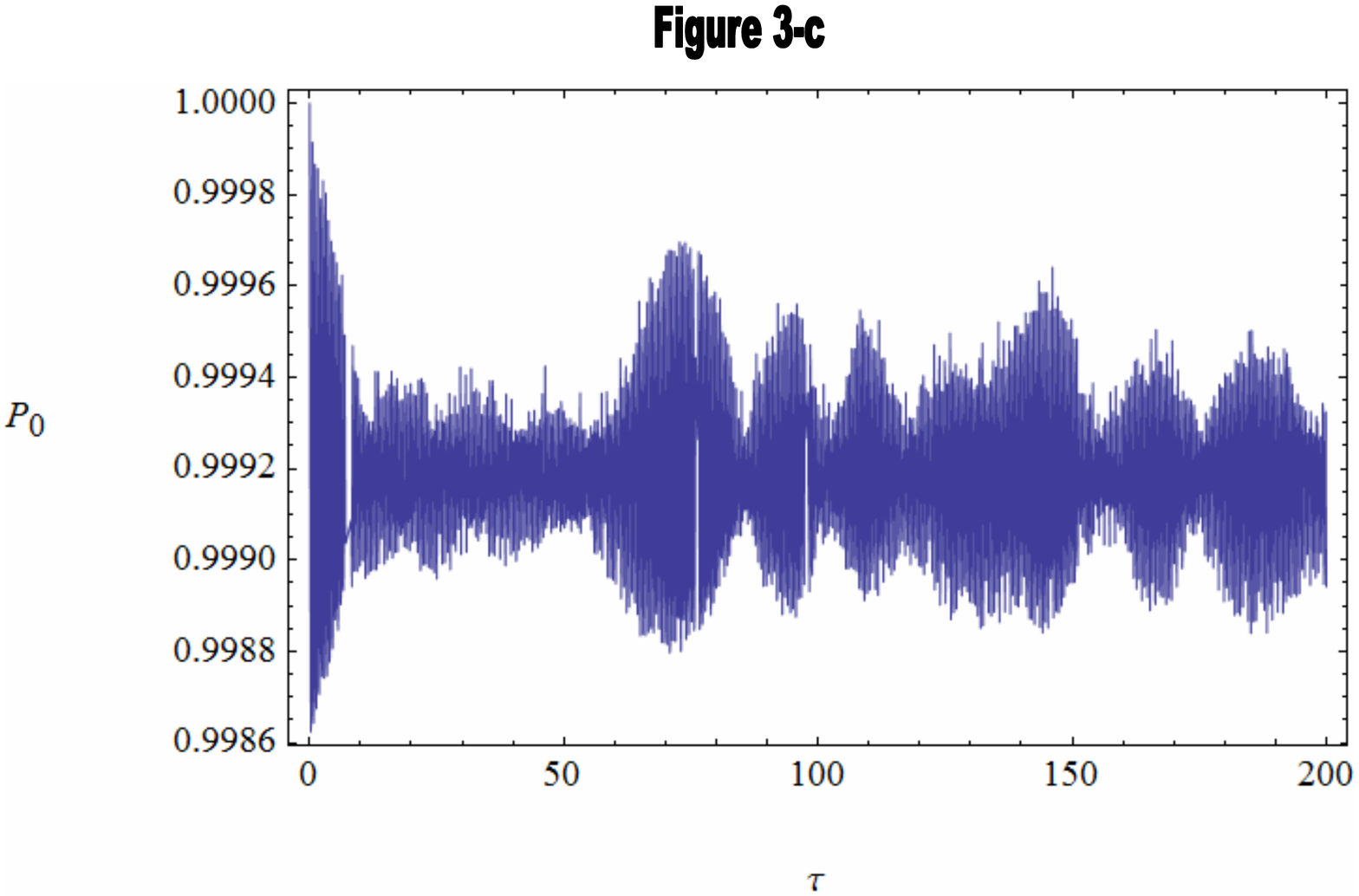}
\includepdf[pages={1}]{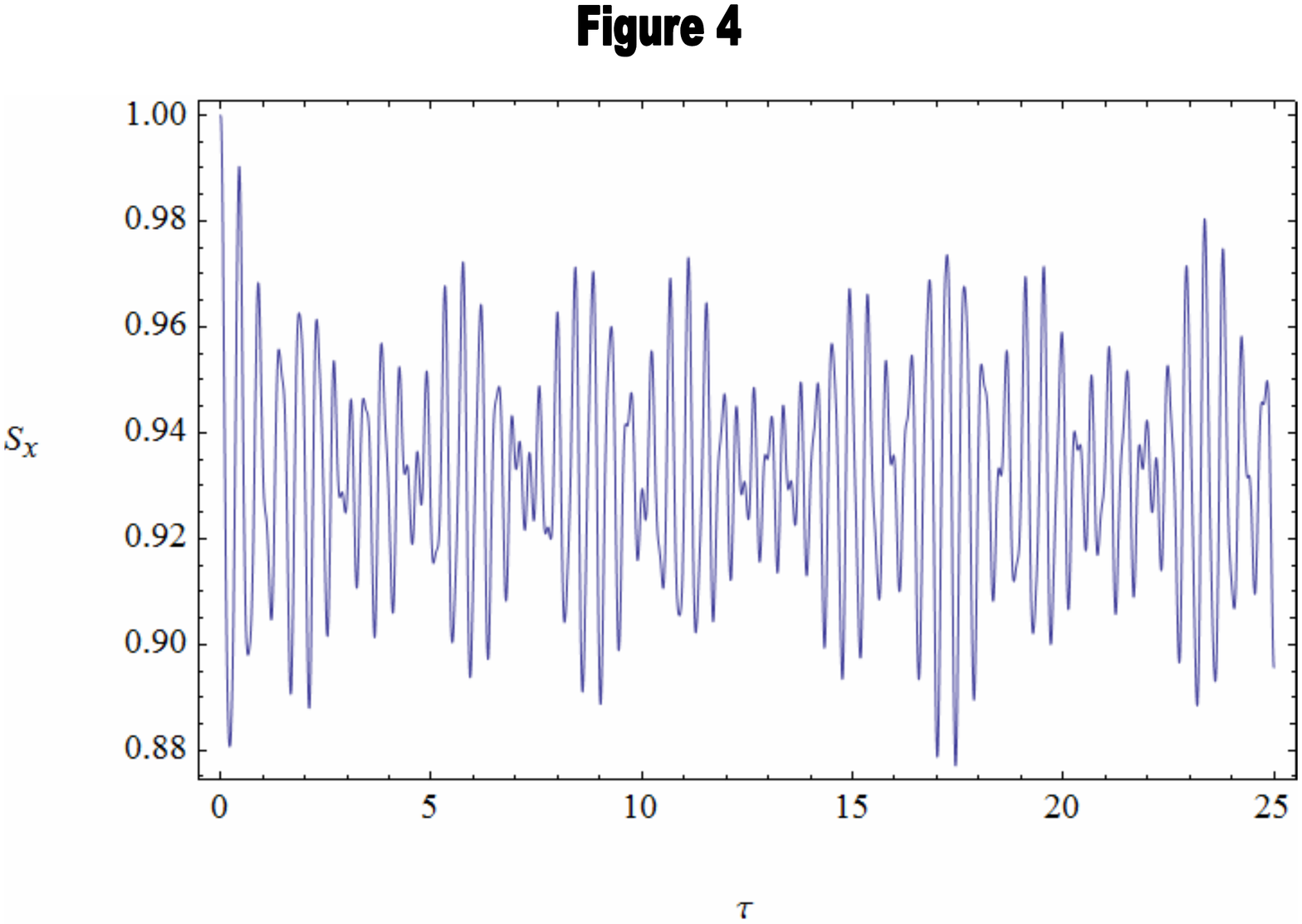}
\includepdf[pages={1}]{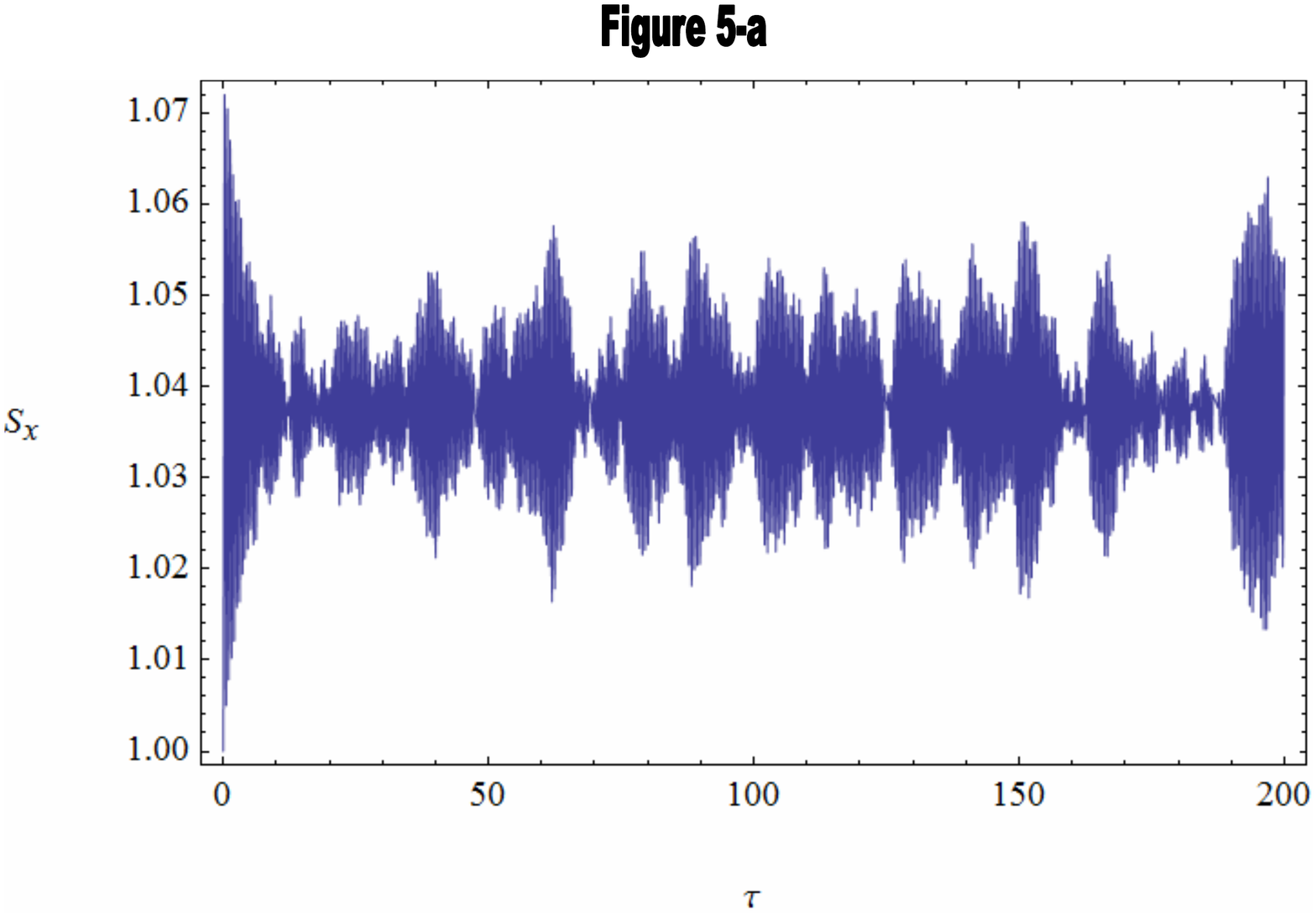}
\includepdf[pages={1}]{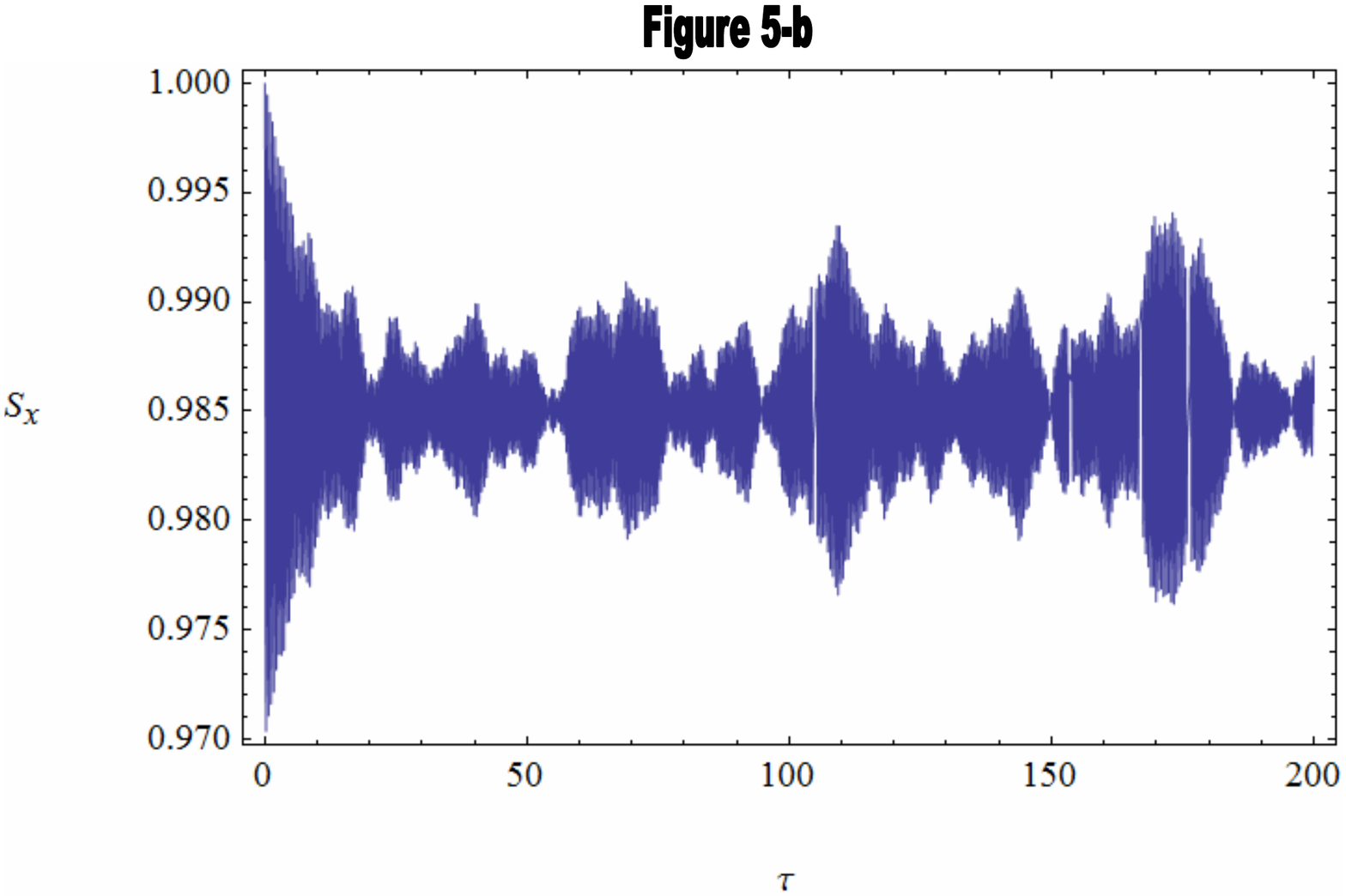}
\includepdf[pages={1}]{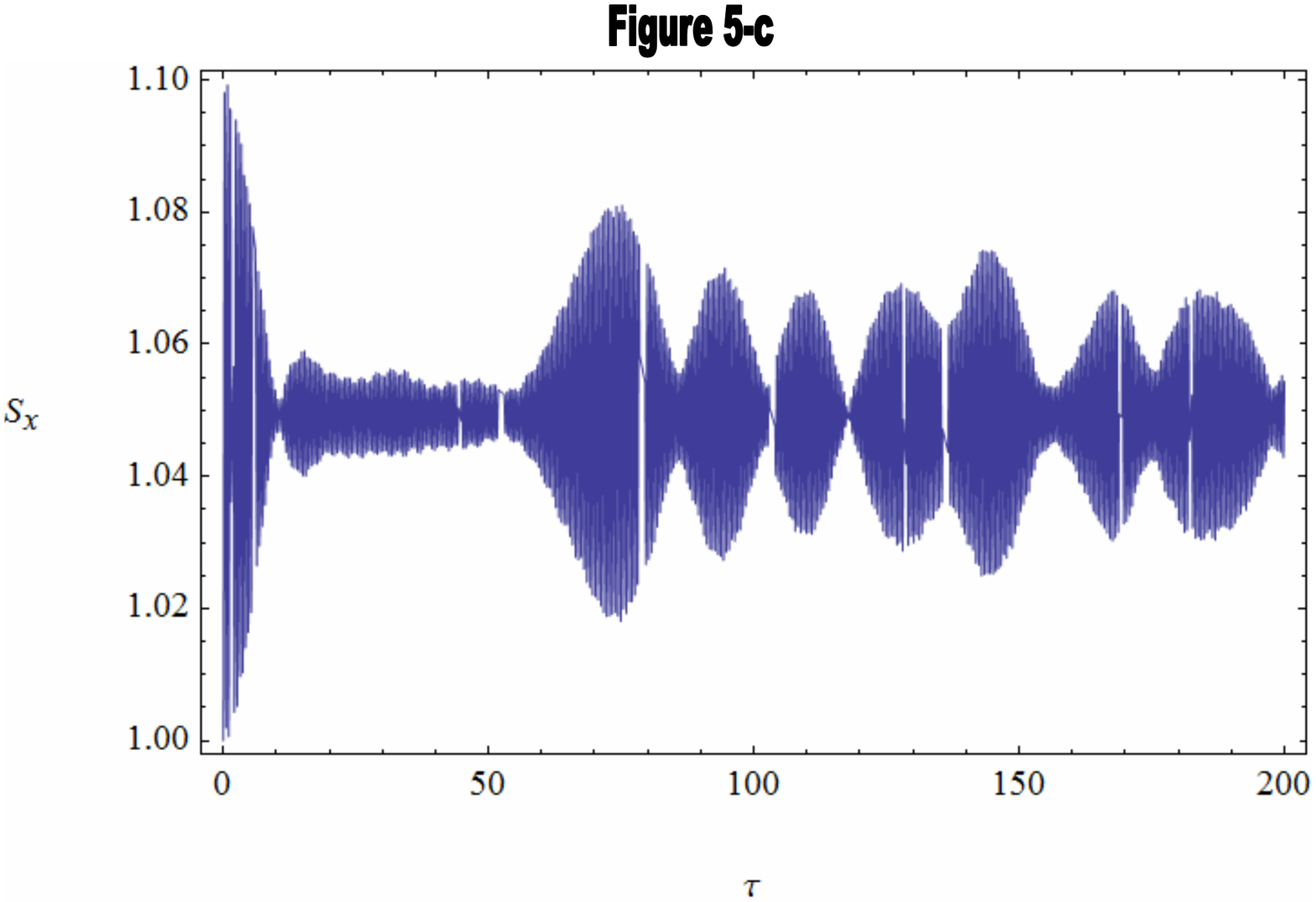}

  \end{document}